\documentclass[floatfix,aps,prb,reprint]{revtex4-1}
\usepackage[latin9]{inputenc}
\usepackage{color}
\usepackage{mathtools}
\usepackage{amsmath}
\usepackage{amssymb}
\usepackage{graphicx}
\usepackage[unicode=true,
 bookmarks=false,
 breaklinks=false,pdfborder={0 0 1},backref=false,colorlinks=true]
 {hyperref}
\hypersetup{
 citecolor=blue,linkcolor=blue}

\makeatletter

\newcommand{\noun}[1]{\textsc{#1}}

\makeatother

\begin{document}
\title{Lasing in a coupled hybrid double quantum dot-resonator system}
\begin{abstract}
We theoretically investigate the possibility of lasing in an electromagnetic
resonator coupled to a voltage-biased hybrid-double-quantum-dot comprised
of a double-quantum-dot tunnel coupled simultaneously to a normal
metal and a superconducting lead. Using a unitary transformation,
we derive a resonator-double-quantum-dot interaction Hamiltonian in
the rotating wave approximation which reveals the fact that lasing
in this system is mediated by electron transitions between Andreev
energy levels in the system's density of states. Moreover, by employing
a markovian master equation incorporating dissipation effects for
the electronic and photonic degrees of freedom, we numerically calculate
the steady-state reduced density-matrix of the system from which we
determine the average photon number and its statistics in the resonator
in various parameter regimes. We find that at some appropriate parameter
configurations, lasing can be considerably enhanced due to the possiblity
of electron tansitions between multiple Andreev levels in the system.
\end{abstract}
\author{S. Mojtaba Tabatabaei}
\email{s.m.taba90@gmail.com}

\affiliation{Faculty of Physics, Shahid Beheshti University, G. C. Evin, Tehran
1983963113, Iran}
\author{Neda Jahangiri}
\affiliation{Faculty of Physics, Shahid Beheshti University, G. C. Evin, Tehran
1983963113, Iran}
\maketitle

\section{\label{I}Introduction}

Laser has been one of the most intriguing optical devices over the
past few decades, and today it has found tremendous number of applications
in different fields of the science and technology. While in conventional
lasers, light amplification is induced by optical pumping of large
number of atoms which are weakly coupled to the cavity mode, the realization
of strongly coupled one-atom-laser in tightly confined cavity modes
is of fundamental interest to the researchers due to their particular
properties and usages\citep{mu1992one,ginzel1993quantum,pellizzari1994preparation,pellizzari1994photon,an1994microlaser,jones1999photon,loffler1997spectral,karlovich2001quantum}.
Following the first experimental achievement in 2003 using a single
Cesium atom strongly coupled to a cavity mode\citep{mckeever2003experimental},
many theoretical proposals and experimental demonstrations were put
forward to explore other realizations of the single atom laser\citep{yoshie2004vacuum,florescu2004theory,aoki2006observation,lougovski2007strongly,florescu2008nonclassical,kim2008one,ritter2010emission}.
Among of these, quantum dot(QD) lasers find a lot of interest because
of their tunability as well as their fabrication advantages. In the
last two decades, based on the cavity quantum electrodynamics, many
micro and nano-cavity lasers with a single-QD have been fabricated\citep{Reithmaier-2004,Reitzenstein,Nomura:09,Nomura-2010,ledentsov2010quantum,liu2011long}. 

Recently, similar effects are demonstrated in the circuit quantum
electrodynamics architecture, where a double-quantum-dot(DQD) or a
superconducting qubit, which is constantly driven into an excited
state by an external electric or magnetic bias, plays the role of
an active media in the electromagnetic resonator formed in a superconducting
transmission line\citep{marthaler2011lasing,jin2011lasing,PhysRevLett.110.066802,liu2015semiconductor,karlewski2016lasing,stehlik2016double,astafiev2007single,PhysRevLett.100.037003,grajcar2008sisyphus,andre2009few}.
In particular, it was theoretically predicted in Ref.{[}\onlinecite{jin2011lasing}{]}
and then experimentally shown in Ref.{[}\onlinecite{liu2015semiconductor}{]},
that a voltage biased DQD can create a lasing state when it is coupled
to a resonator through an electric dipole interaction. By noting that,
it was a DQD coupled to two normal metal leads which is actually considered
in the aforementioned proposals to establish the lasing state in the
resonator, it would be valuable to see whether replacing one of the
normal metal leads by a superconductor can have any implications on
the lasing behavior or not?

In general, by coupling a QD to a normal and also to a superconducting
lead, some new intriguing features arise in the energy configurations
and transport properties of the QD, which are basically due to the
formation of the resonant Andreev reflections at the QD-superconducting
lead interface\citep{sun1999resonant,sun1999photon,PhysRevLett.87.176601,PhysRevB.63.094515,nurbawono2010electron,baranski2013gap,weymann2015andreev}.
Recently, the role of such hybrid single QD structures in creating
a lasing state in an electromagnetic resonator becomes attractive\citep{bruhat2016cavity,PhysRevB.100.085435},
however, the consequences of using a hybrid-DQD in a voltage biased
DQD laser has not been studied, so far. It should be noted that, Bruhat
et.al.\citep{PhysRevB.98.155313}, have recently conducted an experiment
on a hybrid-DQD system coupled to an electromagnetic resonator. They
have not studied the lasing state in their setup, but instead, they
investigated the mechanism of resonator-DQD coupling and observed
a symmetric coupling between the hybrid-DQD and the resonator which
is attributed directly to the roles of superconducting proximity effects
in the DQD.

In this paper, we consider a setup comprising a hybrid-DQD which is
capacitively coupled to an electromagnetic resonator. Using an analytical
treatment of the Hamiltonian of the system and also a numerical simulation
of the full quantum system in the steady state, we analyze the lasing
state in various parameter regimes of this model system. We show that
a lasing state with sub-Poissonian statistics is present in this system
at frequencies equal to the energy differences between Andreev energy
levels in the density of states of the hybrid DQD. There are two Andreev
energy levels corresponding to each energy level of the hybrid DQD
which are arranged in the density of states(DOS) of the system symmetrically
around the chemical potential of the superconducting electrode. Therefore,
depending on the weights of different Andreev levels in the DOS, there
are some specific parameter regimes at which the system shows lasing
due to the electron transitions between the corresponding Andreev
energy levels of hybrid DQD.

The remainder of this paper is organized as follows: In Sec.\ref{II},
the model Hamiltonian is introduced and the underlying formalism of
the paper is set up. Our numerical results are considered in Sec.\ref{ III},
and conclusions are presented in Sec.\ref{IV}.

\section{\label{II}MODEL AND FORMALISM}

\subsection{Model Hamiltonian}

\begin{figure}[t]
\includegraphics[width=8.6cm]{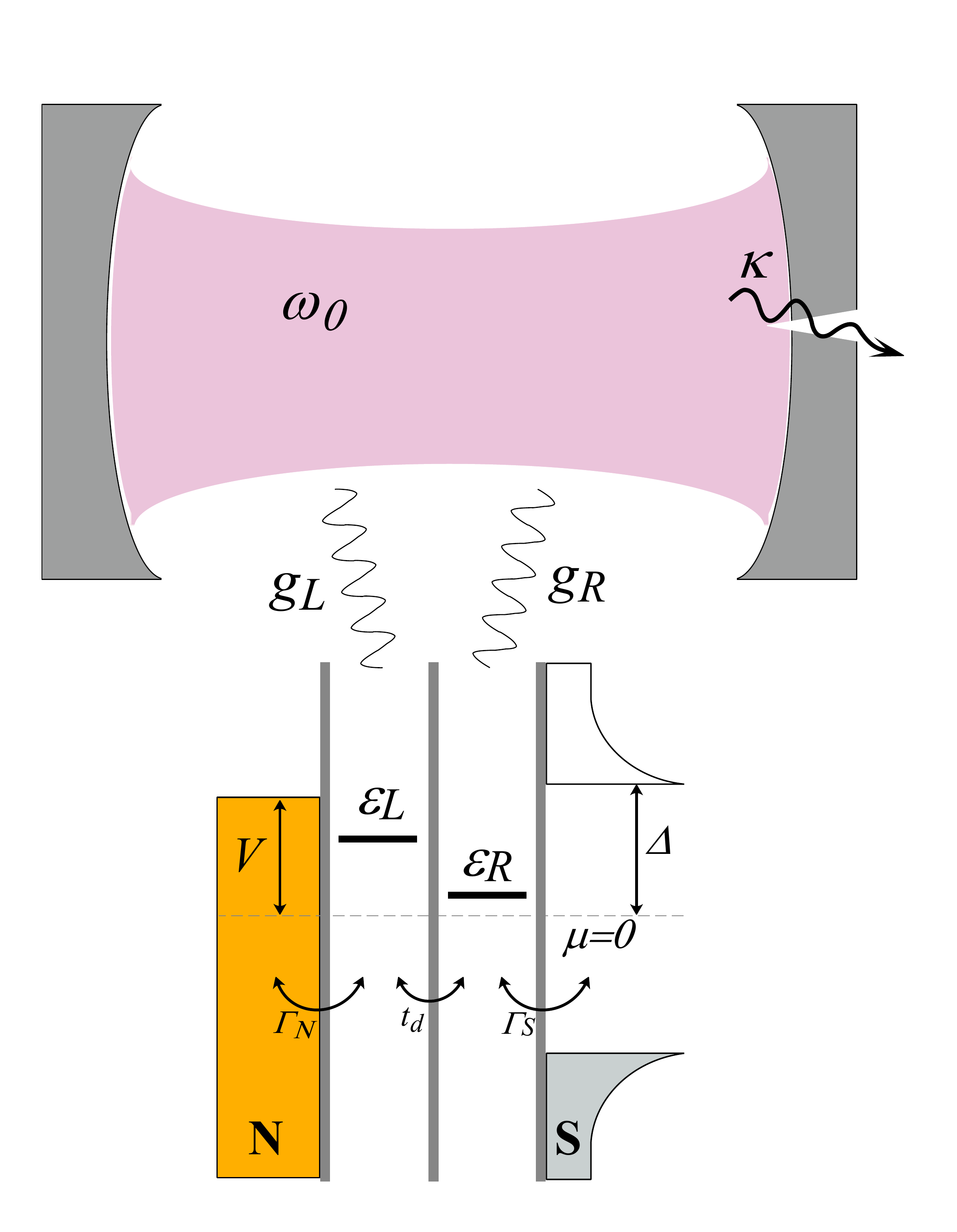} \caption{\label{fig:taba301}A single mode resonator is capacitively coupled
to a voltage-biased hybrid double quantum dot connected to a normal
metal and a superconducting lead.}
\end{figure}
 As it is shown in Fig.\ref{fig:taba301}, our model system consists
of a hybrid DQD, which is capacitively coupled to a single mode electromagnetic
resonator. The total Hamiltonian of the system can be written as:
\begin{equation}
\hat{H}=\hat{H}_{dqd}+\hat{H}_{l}+\hat{H}_{l-d}+\hat{H}_{ph}+\hat{H}_{ph-dqd},
\end{equation}
 where $\hat{H}_{dqd}$ is the Hamiltonian of the DQD given by 
\begin{align}
\hat{H}_{dqd}= & \sum_{\sigma}\varepsilon_{L}\hat{n}_{L,\sigma}+\varepsilon_{R}\hat{n}_{R,\sigma}+t_{d}\sum_{\sigma}(d_{R,\sigma}^{\dagger}d_{L,\sigma}+h.c)\nonumber \\
 & +U_{R}\hat{n}_{R,\uparrow}\hat{n}_{R,\downarrow}+U_{L}\hat{n}_{L,\uparrow}\hat{n}_{L,\downarrow}+U_{LR}\hat{n}_{L}\hat{n}_{R},
\end{align}
where $\hat{n}_{i,\sigma}=\hat{d}_{i,\sigma}^{\dagger}\hat{d}_{i,\sigma}$
is the electron number operator with spin $\sigma=\uparrow,\downarrow$
in the dot $i=L,R$ with energy $\varepsilon_{i}$, and $t_{d}$ is
the hybridization energy between DQDs. Moreover, $U_{L}$ and $U_{R}$
are the onsite Coulomb interaction energies for the left and right
dots, respectively, and $U_{LR}$ is the mutual Coulomb interaction
between the two dots in DQD. Furthermore, $\hat{H}_{l}$ and $\hat{H}_{l-d}$
stand for the Hamiltonian of the leads and tunneling between leads-DQD
which are given by 
\begin{align}
\hat{H}_{l} & =\underset{k,\alpha,\sigma}{\sum}(\varepsilon_{k,\alpha}+\mu_{\alpha})\hat{c}_{k,\alpha,\sigma}^{\dagger}\hat{c}_{k,\alpha,\sigma}\nonumber \\
 & \qquad+\underset{k,\alpha}{\sum}\Delta_{\alpha}(\hat{c}_{k,\alpha,\uparrow}^{\dagger}\hat{c}_{k,\alpha,\downarrow}^{\dagger}+h.c),\\
\hat{H}_{l-d} & =\sum_{k,\alpha,i,\sigma}t_{\alpha,i}(\hat{c}_{k,\alpha,\sigma}^{\dagger}\hat{d}_{i,\sigma}+h.c.),
\end{align}
where $\hat{c}_{\sigma}$ denotes annihilation operator for an electron
with spin $\sigma$ in the single-particle state of the left ($\alpha=N$)
or right lead ($\alpha=S$), characterized by the momentum $k$ with
energy $\varepsilon_{k,\alpha}$, and $\mu_{\alpha}$ is the chemical
potential of the corresponding lead. Moreover, $\Delta_{i}$ is the
pairing energy gap in the respective lead which is given by $\Delta_{N}=0$
for the normal lead and a real positive $\Delta_{S}$ for the superconducting
lead. Furthermore, $t_{\alpha,i}$ is the tunneling energy between
lead $\alpha$ and dot $i$. The left(right) lead is only coupled
to the left(right) dot, thus $t_{N,R}=t_{S,L}=0$. 

Because we are interested in studying the impacts of the presence
of superconducting pairing correlations on the lasing behavior in
the DQD-resonator coupled system, it is reasonable to consider electronic
transitions in the superconducting subgap regime and disregard the
quasiparticle excitations in the continuum region of the superconducting
lead. In practice, this is equal to consider a large superconducting
gap limit, $\Delta_{S}\rightarrow\infty$, in which all energy scales
of the system are smaller than the edges of the superconducting gap.
By using a Green's functions description\citep{trocha2014spin} or
a Schrieffer-Wolf transformation\citep{nigg2015detecting}, it can
be shown that in this so-called infinite-gap approximation, the superconductor
lead is decoupled from right dot, leaving an effective pairing potential
$\Gamma_{S}(\hat{d}_{R,\uparrow}^{\dagger}\hat{d}_{R,\downarrow}^{\dagger}+h.c)$
in the Hamiltonian of DQD, where $\Gamma_{S}=2\pi|t_{S,R}|^{2}\rho_{0}^{S}$
is the electron tunneling rate between DQD and superconducting lead
in the wide-band approximation and $\rho_{0}^{S}$ is the lead's density
of states in its normal state. Thus, we can rewrite the Hamiltonian
of the DQD as 
\begin{gather}
\hat{H}_{dqd}^{SC}=\hat{H}_{dqd}+\Gamma_{S}(\hat{d}_{R,\uparrow}^{\dagger}\hat{d}_{R,\downarrow}^{\dagger}+h.c).
\end{gather}

The Hamiltonian of the single mode resonator is given by $\hat{H}_{ph}=\hbar\omega_{0}(\hat{a}^{\dagger}\hat{a}+\frac{1}{2}),$
where $\hat{a}$ is the bosonic annihilation operator of the resonator
mode with frequency $\omega_{0}$. The coupling of the DQD to the
resonator is modeled by a capacitive interaction between electrons
in the DQD and the electric field in the resonator and its Hamiltonian
is given by 
\begin{equation}
\hat{H}_{ph-dqd}=-\sum_{i,\sigma}g_{i}\hat{n}_{i,\sigma}(\hat{a}+\hat{a}^{\dagger}),
\end{equation}
 where $g_{i}$ is the coupling strength between the dot $i$ and
the resonator mode.

\subsection{Master equation description}

The coupled DQD-resonator system can be driven into a non-equilibrium
state by applying a finite bias voltage $\mu_{N}=eV_{b}$ on the normal
electrode where $e$ is the charge of electron. Following the standard
recipe for deriving the master equation, we consider the DQD-resonator
subsystem with the Hamiltonian $\hat{H}_{S}=\hat{H}_{dqd}^{SC}+\hat{H}_{ph}+\hat{H}_{ph-dqd}$,
as an open system which we seek for its dynamics when it is weakly
coupled to the normal metal electrode and a photon bath. Then, the
dynamics of the system can be described by using a master equation
of the reduced density matrix, $\rho$, of the coupled DQD-resonator
system which in the Markovian approximation is given by\citep{PhysRevB.97.035305,PhysRevB.100.035412}
\begin{align}
\frac{d}{dt}\hat{\rho}(t)=- & \frac{i}{\hbar}\big[\hat{H}_{S},\hat{\rho}(t)\big]+\mathcal{L}_{ph}\hat{\rho}(t)+\mathcal{L}_{N}\hat{\rho}(t),\label{eq:master_eq}
\end{align}
where $\mathcal{L}_{ph}$ and $\mathcal{L}_{N}$ are, repectively,
the Lindblad superoperators describing the dissipation of photons,
and the electron tunneling between the QD and the normal metal electrode.
While for a general description of the dynamics of the DQD-resonator
system, a precise microscopic derivation of the above Lindblad superoperators
is necessary, but as we are interested here to investigate the possibility
of lasing in the hybrid DQD-resonator system, we will focus on the
situations where the system is in low-temperatures and also there
is a large bias voltage applied on the normal metal lead. These assumptions
introduce great simplifications in our calculations and allow us to
represent the action of the above Lindblad superoperators on the reduced
density matrix of DQD-resonator by the following forms:
\begin{equation}
\mathcal{L}_{ph}\hat{\rho}(t)=\kappa\big[\hat{a}\hat{\rho}(t)\hat{a}^{\dagger}-\frac{1}{2}\big\{\hat{a}^{\dagger}\hat{a},\hat{\rho}(t)\big\}\big],\label{eq:eq8}
\end{equation}
and 
\begin{equation}
\mathcal{L}_{N}\hat{\rho}(t)=\Gamma_{N}\sum_{i,\sigma}\big[\hat{C}_{i\sigma}\hat{\rho}(t)\hat{C}_{i\sigma}^{\dagger}-\frac{1}{2}\big\{\hat{C}_{i\sigma}^{\dagger}\hat{C}_{i\sigma},\hat{\rho}(t)\big\}\big],\label{eq:eq9}
\end{equation}
where $\kappa$ is the decay rate of the photons in the resonator,
$\Gamma_{N}=2\pi|t_{N,L}|^{2}\rho_{0}^{N}$ is the electronic tunneling
rate to the normal metal lead, and $\hat{C}_{i\sigma}=\hat{d}_{i,\sigma}^{\dagger}(\hat{d}_{i,\sigma})$,
for positive(negative) bias voltages. We emphasize that if we had
not considered the infinite-gap approximation for the superconducting
lead, it would have been necessary in Eq.(\ref{eq:master_eq}) to
consider the effect of coupling to the superconducting lead by introducing
its corresponding Lindblad superoperator\citep{PhysRevB.87.155439}.
We also note that the above formalism best describes the coupled DQD-resonator
system when the coupling between DQD and resonator is weak. For a
general treatment in the presence of strong coupling between the DQD
and the resonator we can refer to the Ref.{[}\onlinecite{PhysRevB.100.035412}{]}.

By solving Eq.(\ref{eq:master_eq}) for $d\hat{\rho}(t)/dt=0$, we
obtain the reduced density matrix of the coupled DQD-resonator system
in the steady state, from which we can calculate the average value
of every observable of the system using the relation $\bigl\langle\hat{O}(t)\bigr\rangle=\text{Tr}\bigl[\hat{O}\hat{\rho}(t)\bigr]$,
where $\text{Tr}[...]$ is the trace with respect to all degrees of
freedom in the system. Of interest to us here are the average photon
number, $n_{photon}=\bigl\langle a^{\dagger}a\bigr\rangle$, and the
Fano factor of the photons where the latter can be calculated by the
relation $F=(\bigl\langle a^{\dagger}aa^{\dagger}a\bigr\rangle-\bigl\langle a^{\dagger}a\bigr\rangle^{2})/\bigl\langle a^{\dagger}a\bigr\rangle$.
The Fano factor can be referred to as a means to distinguish between
the photon bunching($F>1$) or anti-bunching($F<1$) regimes and describes
whether the photons inside the resonator have either sub-Poissonian($F<1$),
Poissonian($F=1$) or super-Poissonian($F>1$) statistics.\citep{PhysRevB.93.235450}

With the knowledge of the reduced density matrix of the system, we
can also calculate the distribution of the photons number in the resonator,
which is given by the diagonal elements of $\text{Tr}_{QD}\bigl[\hat{\rho}(t)\bigr]$
in the occupation number basis, where $\text{Tr}_{QD}[...]$ is the
trace with respect to the degrees of freedom of the QD. 

\subsection{Rotating-wave-approximation}

Despite its simplicity, Eq.(\ref{eq:master_eq}) cannot be solved
analytically and it is necessary to solve it numerically. Before presenting
our numerical results, it is instructive to examine the Hamiltonian
of the system in a rather different representation which is more familiar
in the context of the quantum optics literature\citep{breuer2002theory}.
By transforming to a representation in which the Hamiltonians $\hat{H}_{dqd}^{SC}$
and $\hat{H}_{ph}$ are diagonal, we can find a rotating-wave-approximation(RWA)
description of $\hat{H}_{ph-dqd}$ by which a clear interpretation
of the lasing mechanism can be obtained easily. However, the simultaneous
presence of pairing($\Gamma_{S})$ and tunneling($t_{d}$) terms in
$\hat{H}_{dqd}^{SC}$, makes its analytical diagonalization practically
impossible. Nevertheless, we can use perturbation theory to obtain
a unitary transformation to the lowest order in $\Gamma_{S}$, which
can be used to obtain an approximating expression for diagonalized
$\hat{H}_{dqd}^{SC}$. 

We relegate the details of deriving such unitary transformation to
the Appendix\ref{sec:Diagonalization-of-}, and here we solely present
the results. Using a unitary transformation $\hat{U}$, which its
explicit representation is given in Eq.(\ref{eq:eq17-1}), the Hamiltonian
$\hat{H}_{dqd}^{SC}$ becomes diagonalized as 
\begin{equation}
\hat{\widetilde{H}}_{dqd}^{SC}=\hat{U}^{\dagger}\hat{H}_{dqd}^{SC}\hat{U}=\sum_{i,\sigma}E_{i}\hat{\gamma}_{i,\sigma}^{\dagger}\hat{\gamma}_{i,\sigma}+\mathcal{O}(\Gamma_{S}^{2}),
\end{equation}
where $E_{1}(E_{2})=(\varepsilon_{L}+\varepsilon_{R})/2\pm[((\varepsilon_{L}-\varepsilon_{R})/2)^{2}+t_{d}^{2}]^{1/2}$
and the fermionic operators $\hat{\gamma}_{i,\sigma}$ with $i=1,2$
and $\sigma=\uparrow,\downarrow$, are related to the $\hat{d}_{i,\sigma}$
operators through Eq.(\ref{eq:eq17-1}). By applying the same unitary
transformation on the DQD-resonator coupling Hamiltonian, $\hat{H}_{ph-dqd}$,
and after some algebra, we reach to the following expression for the
transformed Hamiltonian $\hat{\widetilde{H}}_{ph-dqd}$, in the rotating-wave-approximation:
\begin{gather}
\hat{\widetilde{H}}_{ph-dqd}^{\textrm{RWA}}=uv\sum_{\sigma}\hat{a}\left\{ (g_{L}-g_{R})\left[\hat{\gamma}_{1,\sigma}^{\dagger}\hat{\gamma}_{2,\sigma}\right.\right.\hspace{1cm}\nonumber \\
\left.-\Gamma_{S}(\frac{u^{2}}{2E_{2}}+\frac{v^{2}}{2E_{1}})\hat{\gamma}_{1,\sigma}^{\dagger}\hat{\gamma}_{2,-\sigma}^{\dagger}\right]\nonumber \\
\left.-(g_{L}+g_{R})\Gamma_{S}\frac{1}{E_{1}+E_{2}}\hat{\gamma}_{1,\sigma}^{\dagger}\hat{\gamma}_{2,-\sigma}^{\dagger}+\mathcal{O}(\Gamma_{S}^{2})\right\} +h.c,\label{eq:RWA}
\end{gather}
where $u(v)=\frac{1}{\sqrt{2}}[1\pm\frac{\varepsilon_{L}-\varepsilon_{R}}{[(\varepsilon_{L}-\varepsilon_{R})^{2}+4t_{d}^{2}]^{1/2}}]^{1/2}.$
Note that for $\Gamma_{S}=0$, the above equation is reduced to the
usual electric dipole coupling between a DQD and a resonator. We emphasis
that, to the linear order of $\Gamma_{S}$, in addition to the term
proportional to $g_{L}-g_{R}$ which is usual for the electric dipole
coupling of a DQD to a resonator, there appears a new term in Eq.(\ref{eq:RWA})
which is proportional to $g_{L}+g_{R}$. In fact, this new intriguing
DQD-resonator coupling term is particularly due to the coupling of
the superconducting lead to the DQD and its effects have been observed
only very recently in Ref.{[}\onlinecite{PhysRevB.98.155313}{]}.
Another interesting thing which we can understand from Eq.(\ref{eq:RWA})
is that a photon creation is accompanied with a new electronic transition
in the DQD represented by the operator $\hat{\gamma}_{1,\sigma}^{\dagger}\hat{\gamma}_{2,-\sigma}^{\dagger}$
which is related to the Andreev reflections in DQD due to the coupling
to the superconducting lead. It should be noted that Eq.(\ref{eq:RWA})
is obtained perturbatively to the first order of $\Gamma_{S}$, and
it should not be used for a quantitative explanation of the lasing
in the DQD-resonator coupled system in the general case.

\section{\label{ III}Results and Discussions}

In order to solve Eq.(\ref{eq:master_eq}) numerically, we need to
express its matrix representation in an appropriate basis. We can
express the operators in Eq.(\ref{eq:master_eq}) in the basis spanned
by the states $\left|n_{L\uparrow}\right\rangle \otimes\left|n_{L\downarrow}\right\rangle \otimes\left|n_{R\uparrow}\right\rangle \otimes\left|n_{R\downarrow}\right\rangle \otimes\left|n_{p}\right\rangle $,
where $n_{i,\sigma}=0,1$ is the electron occupation number and $n_{p}=0,1,2,...$,
represents the number of photons in the resonator. In practice, we
need to truncate the maximum value of photon numbers in the resonator
to a sufficient value $n_{p,max}$, which should be taken large enough
to assure the convergence of the numerical calculations. In our calculations,
$n_{p,max}=70$, is found to be a sufficient cutoff number. It is
noteworthy that to calculate the steady state reduced density matrix,
we need to solve a set of $(2^{4}\times(n_{p,max}+1))^{2}$ equations,
which is a highly time and memory consuming process. The uniqueness
of the calculated reduced density matrix is guaranteed by using the
normalization condition $\text{Tr}\left[\hat{\rho}(t)\right]=1$.
Our numerical calculations were performed by utilizing \noun{QuTiP}
package\citep{johansson2013qutip}.

Since there are no previous experiments addressing a similar model
as what we considered in this work, we will take the values of our
model parameters on the order of experimentally accessible values
which are reported in some previous experimental works\citep{doh2008andreev,PhysRevLett.113.036801,PhysRevB.98.155313}.
So, we take the energy of the left dot level and the interdot tunneling
energy as $\varepsilon_{L}=10\mu eV$ and $t_{d}=5\mu eV$. Also,
we set $\Gamma_{N}=0.1\sim1\mu eV$ and $\Gamma_{S}=1\sim5\mu eV$.
The chemical potential of the superconducting electrode is taken as
the energy reference, $\mu_{S}=0$, and a large positive bias voltage
is applied directly on the normal electrode, $\mu_{N}=eV_{b}$. Moreover,
we set the resonator damping to $\kappa=10^{-3}\mu eV$ and also the
magnitude of the DQD-resonator coupling to $\left|g_{L}\right|=\left|g_{R}\right|=g_{0}=6.62\times10^{-2}\mu eV$.
To simplify our discussion and for more clarity, we momentarily disregard
the electron-electron interactions in the DQD and demonstrate the
presence of the lasing state in the noninteracting case. We shall
come back to the issue of the presence of nonzero electron-electron
interactions in the DQD at the final stage of this section.

\begin{figure}
\includegraphics[width=8.6cm]{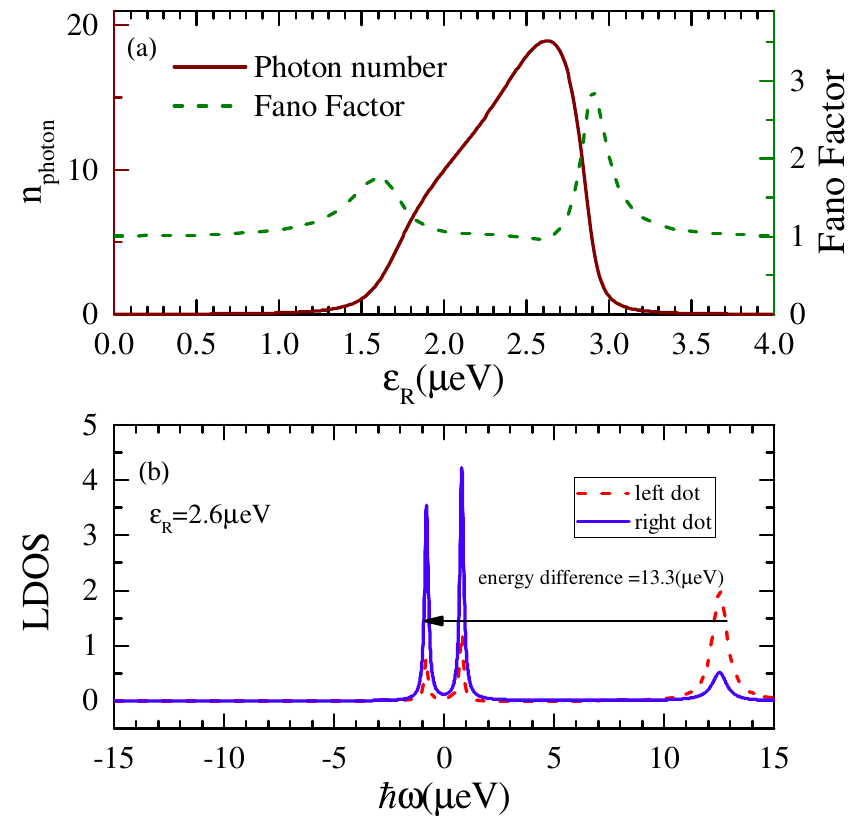} \caption{\label{fig:1}(a) Average photon number(solid brown) and Fano factor(dashed
green) as functions of the energy level of right dot, $\varepsilon_{R}$.
(b) Local density of states of the DQD for $\varepsilon_{R}=2.6\mu eV$.
Other parameter are $\varepsilon_{L}=10\mu eV$, $U_{L}=U_{R}=U_{LR}=0$,
$\Gamma_{N}=0.1\mu eV$, $\Gamma_{S}=1\mu eV$, $\mu_{S}=0$, $\hbar\omega_{0}=13.3\mu eV$,
$g_{L}=-g_{R}=g_{0}=6.62\times10^{-2}\mu eV$ and $\kappa=10^{-3}\mu eV$. }
\end{figure}
In Fig.\ref{fig:1}(a), we fix the frequency of the resonator to $\hbar\omega_{0}=13.3\mu eV$,
and plot the average photon number and the Fano factor of the photons
in the resonator as a function of the energy level of right dot, $\varepsilon_{R}$.
We see that the average photon number in the resonator shows an asymmetric
peak with maximum photon number about $n_{photon}=20$, around $\varepsilon_{R}=2.6\mu eV$,
accompanied with a Fano factor slightly lower than $F=1$, corresponding
to a sub-Poissonian distribution for the photons in the resonator.
A rigorous way to interpret why the lasing happens at these particular
gate voltage and resonator frequency and also why the photon peak
is asymmetric, is to look at the local density of states(LDOS) of
the DQD when it is in equilibrium state and isolated from the resonator
and the normal lead. The LDOS can be easily obtained from $-\textrm{Im}[G_{i}^{R}(\omega)]$,
where $G_{i}^{R}(\omega)$ is the retarded Green's function of the
dot $i$ which can be calculated by using the Lehmann representation(See
Appendix\ref{sec:Lehmann})\citep{bruus2004many}. In general, by
coupling to a superconducting lead, the peaks in the LDOS of a QD
with energies around the chemical potential of the superconducting
lead, will be split into two Andreev reflection subgap levels with
equal but opposite sign energies\citep{baranski2013gap}. On the other
hand, since we have assumed a large positive bias voltage on the left
lead, we can expect that the direction of electron tunnelings should
be from the left to the right dot. If the energy levels of the two
dots are in resonance, we will end up with an electric current originated
from resonant Andreev reflections at the interface of the right dot
and the superconducting lead\citep{sun1999resonant,PhysRevB.81.075404}.
However, when the energy levels of the two dots are misaligned, electron
tunnelings from left to right dots are assisted by some photon excitation
or absorption in the resonator, the frequency of which is equal to
the energy difference between the two corresponding LDOS peaks of
the left and right dot. 

We have plotted the LDOS of the left and right dots in Fig.\ref{fig:1}(b),
for the gate voltage at which the the photon number in the resonator
is maximized. We see that the LDOS of left dot, has a large peak around
$\hbar\omega\approx12.3\mu eV$, while the LDOS of the right dot has
two main peaks at energies $\hbar\omega\approx\pm1\mu eV$. It can
be deduced that the photon number peak in Fig.\ref{fig:1}(a), which
is for resonator frequency $\hbar\omega_{0}=13.3\mu eV$, is because
of electron transitions between the two peaks in the LDOS of DQD by
an arrow in Fig.\ref{fig:1}(b). Interestingly, we see that there
is another possible electron transition with energy difference $\hbar\omega\approx11.3\mu eV$,
which we anticipate we can see its lasing effect by tuning the frequency
of the resonator to the appropriate frequency. The origin of the asymmetry
in the photon peak in Fig.\ref{fig:1}, can also be deduced from the
LDOS by noting that the weight of the Andreev subgap peaks in the
LDOS is dependent on the values of the gate voltages. 

\begin{figure}
\includegraphics[width=8.6cm]{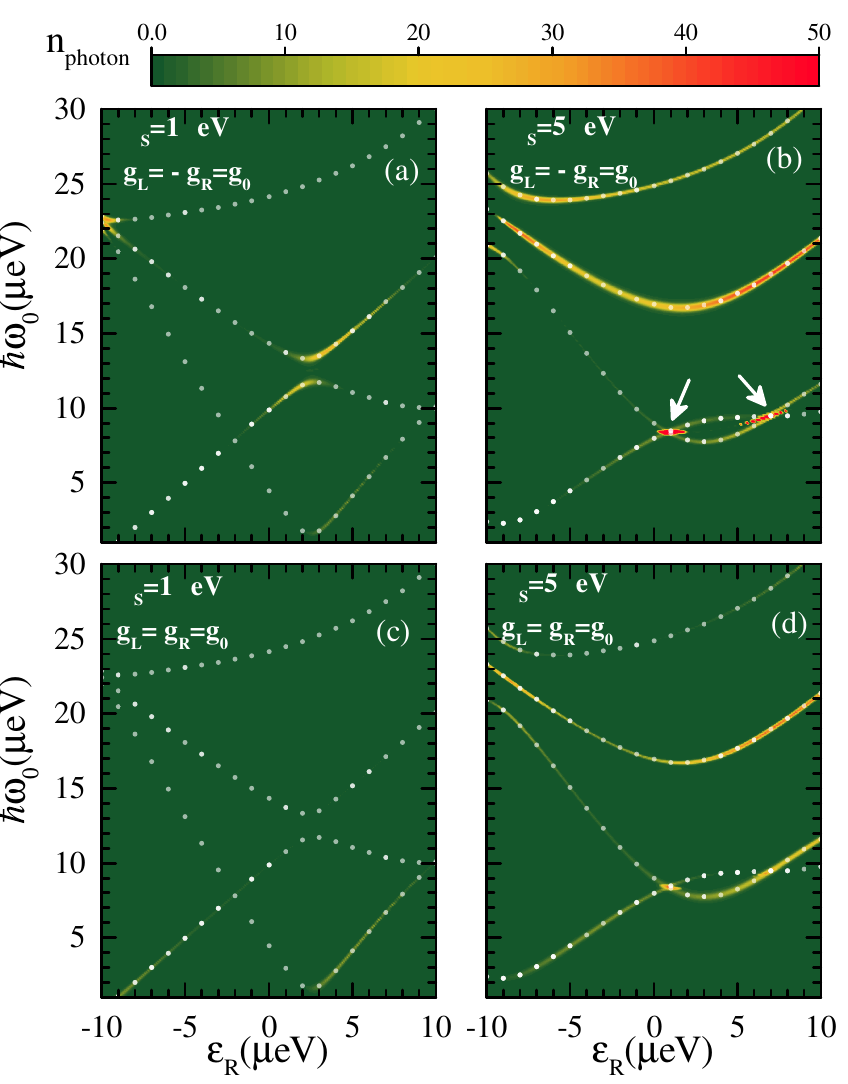}\caption{\label{fig:3}(a)-(d) Average number of photons in the resonator as
a function of resonator frequency, $\omega_{0}$, and the energy level
of right dot, $\varepsilon_{R}$, for various values of $\Gamma_{S}$
and $g_{L}$ and $g_{R}$. (a) $\Gamma_{S}=1\mu eV$ and $g_{L}=-g_{R}=g_{0},$
(b) $\Gamma_{S}=5\mu eV$ and $g_{L}=-g_{R}=g_{0},$ (c) $\Gamma_{S}=1\mu eV$
and $g_{L}=g_{R}=g_{0},$ (d) $\Gamma_{S}=5\mu eV$ and $g_{L}=g_{R}=g_{0}$.
White circles show the energy difference of various peaks in the LDOS
at the respective $\varepsilon_{R}$. Other parameters are as in Fig.\ref{fig:1}.}
\end{figure}
From the above discussion, it becomes clear that because of the presence
of the superconducting lead, there are of course more than one possible
lasing frequencies in our model system corresponding to each parameter
configuration of the DQD. To see this, we have plotted false color
plots of the average photon number in the resonator as a function
of $\varepsilon_{R}$ and $\omega_{0}$ in Fig.\ref{fig:3}. Additionally,
we have also calculated energy differences between various LDOS peaks
of the system at the different gate voltages, which are shown in Fig.\ref{fig:3}
by white circles. In Fig.\ref{fig:3}(a), we consider the same parameter
configuration as in Fig.\ref{fig:1}. We see that lasing happens at
various gate voltages as well as in different resonator frequencies.
Moreover, as it is expected, the resonator frequencies at which we
can see nonzero lasing are exactly equal to the energy differences
between various LDOS peaks. It is interesting to note that the regions
with nonzero lasing do not follow regular pattern which is mainly
because of the fact that positions and heights of the LDOS peaks are
non-linear functions of the values of $t_{d}$, $\varepsilon_{R}$,
$\varepsilon_{L}$ and $\Gamma_{S}$. Alternatively, this non-linearity
can also be seen in the approximate expression which we obtained for
the DQD-resonator coupling in Eq.(\ref{eq:RWA}). 

Next, in Fig.\ref{fig:3}(b), we investigate the effect of increasing
the coupling energy between the right dot and the superconducting
lead by setting $\Gamma_{S}=5\mu eV$. Because of increasing the value
of $\Gamma_{S}$, the LDOS peaks are renormalized and therefore their
energy differences are also changed. Accordingly, the profile of the
lasing which should follow these energy differences is changed as
well. An important feature in Fig.\ref{fig:3}(b) is the presence
of two points, which are marked by two white arrows. We see that at
these points, two branches with nonzero lasing are crossing and, very
intriguingly, as a result of this branch crossing the photon number
in the resonator is considerably increased. Actually, the crossing
of two energy difference branches means that there are two distinct
possible electron transitions in the LDOS with exactly the same energy
difference. Hence, we can deduce that the sudden increasing of the
photon number at these points is due to multiple photon excitation
due to the electron transitions from different pairs of peaks in the
LDOS. Another feature in Fig.\ref{fig:3}(b) is that, in the regions
around the crossing points, some nonzero lasing occurs at frequencies
which are not associated to any energy difference branches. We speculate
that these nonzero lasing points are originated from two-level lasing
mechanism corresponding to the electron transitions from the two nearby
branches\citep{markus2003simultaneous}.

So far, in Figs.\ref{fig:1} and \ref{fig:3}(a) and (b), we have
investigated the properties of the lasing state for the situation
where the DQD-resonator coupling is asymmetric where $g_{L}=-g_{R}=g_{0}$.
However, as we have shown by using some approximative calculations
in Sec.\ref{II}, a spectacular character of the coupling a hybrid
DQD to the resonator, is the presence of new coupling mechanism which
is dependent on the sum of the coupling strength of each dot to the
resonator, $g_{L}+g_{R}$, and therefore, one can obtain lasing in
this system even for symmetric DQD-resonator coupling as $g_{L}=g_{R}$.
Thus, it is interesting to study the presence of lasing in such symmetric
couplings. We have plotted in Figs.\ref{fig:3}(c) and (d), the same
plots as in (a) and (b), except that here we consider the case of
$g_{L}=g_{R}=g_{0}$. Here, because the energy configurations of the
DQD are not changed, the profile of the energy differences are not
altered. However, we see that there are still some nonzero lasing
states available for the system in this symmetric coupling configuration.
We emphasize that in the case where the DQD is coupled to two normal
metals, the DQD-resonator coupling is described only by a term proportional
to $g_{L}-g_{R}$, and one could expect that a symmetric DQD-resonator
coupling as $g_{L}=g_{R}$, would result in a completely vanishing
DQD-resonator coupling and therefore no lasing can happen in such
situations. 

\begin{figure}
\includegraphics[width=8.6cm]{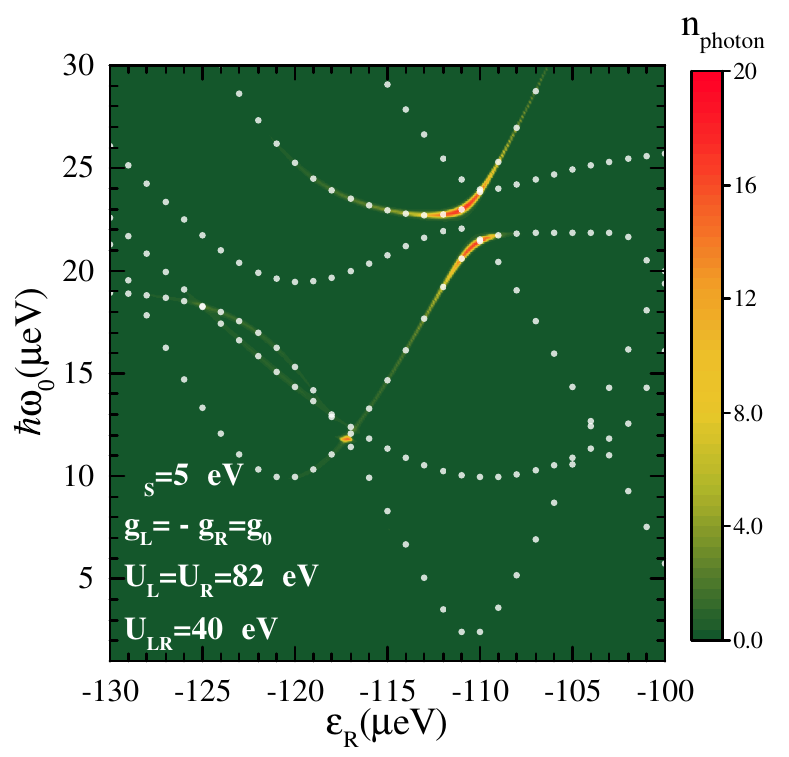}\caption{\label{fig:4}Average number of photons in the resonator as a function
of resonator frequency, $\omega_{0}$, and the energy level of right
dot, $\varepsilon_{R}$, for the presence of finite electron-electron
interactions in DQD. We set $U_{L}=U_{R}=82\mu eV$ and $U_{LR}=40\mu eV$
and other parameters are $\varepsilon_{L}=-110\mu eV$, $\Gamma_{N}=1\mu eV$,
$\Gamma_{S}=5\mu eV$, $\mu_{S}=0$, $g_{L}=-g_{R}=g_{0}$ and $\kappa=10^{-3}\mu eV$. }
\end{figure}
Having realized the presence of the lasing in the coupled hybrid DQD-resonator
system as well as its origins and some of its properties, we now study
the effect of nonzero electron-electron interactions on the lasing
state in our model system. Although, it seems that the presence of
finite electron-electron interactions in the hybrid-DQD could suppress
the superconducting proximity correlations in the DQD and accordingly
could reduce the lasing state, however, previous theoretical and experimental
analysis have shown that indeed the electronic transport through a
hybrid QD, mainly depends on three energy scales, namely $U/\Gamma_{S}$,
$\Gamma_{N}/\Gamma_{S}$ and $t_{d}/\Gamma_{S}$\citep{doh2008andreev,PhysRevB.81.075404,deacon2010tunneling}.
So, even for large $U$ values, we can expect nonzero conductance
through DQD and therefore the presence of lasing in this case is also
of no surprise. As a representative example, in Fig.\ref{fig:4},
we investigate the presence of lasing for a parameter configuration
as in Fig.\ref{fig:3}(b) except that here we consider a large electron-electron
interaction in the DQD by setting $U_{L}=U_{R}=82\mu eV$ and $U_{LR}=40\mu eV$,
which are roughly similar to the values given in Ref.{[}\onlinecite{PhysRevB.98.155313}{]}.
As a result of electron-electron interaction, the energy difference
branches are split into many branches (some of them are not shown
in Fig.\ref{fig:4}) and among them, at some particular regions, we
can see some nonzero lasing states in various resonator frequencies.
Note the large negative gate voltages $\varepsilon_{L}$ and $\varepsilon_{R}$
applied on the DQD in the Fig.\ref{fig:4}, which are necessary to
compensate the large Coulomb interactions in the DQD. Also, note that
the energy difference branches are obtained from the energy differences
of the roots of the retarded Green's function, which is calculated
for the full interacting system by using the Lehmann representation.

\section{\label{IV}Conclusions}

In this paper, we studied the possibility of lasing in a single mode
electromagnetic resonator which is capacitively coupled to a hybrid
DQD. We found that lasing in this system is mediated by electron tunneling
between various peaks in the LDOS of the DQD which are induced by
Andreev reflections at the DQD-superconducting lead's interface. Because
of this particular lasing mechanism, we showed that the average photon
number in the resonator in the lasing state can be further enhanced
by multiple electron tunnelings between two different pairs of Andreev
reflection peaks with the same energy difference in the LDOS of the
DQD. Also, we showed that this system could also exhibit ``two-state
lasing'' which is due to the electron transitions between different
Andreev levels in the hybrid DQD with different energy differences. 
\begin{acknowledgments}
We are grateful to Farshad Ebrahimi for useful discussions. Numerical
calculations were performed by using high performance computational
facilities of Shahid Beheshti University (SARMAD).
\end{acknowledgments}

\appendix

\section{\label{sec:Diagonalization-of-}Diagonalization of $\hat{H}_{dqd}^{SC}$ }

Here, we present the details of steps needed for diagonalizing the
Hamiltonian $\hat{H}_{dqd}^{SC}$. In these calculations we disregard
the Coulomb interaction in the DQD to simplify our presentation. We
start with the matrix representation 
\begin{equation}
\hat{H}_{dqd}^{SC}=\hat{\Psi}^{\dagger}(h_{0}+\Gamma_{S}h_{1})\hat{\Psi},\label{eq:eq12}
\end{equation}
where $\hat{\Psi}^{\dagger}=(\hat{d}_{1,\uparrow}^{\dagger},\hat{d}_{1,\downarrow},\hat{d}_{2,\uparrow}^{\dagger},\hat{d}_{2,\downarrow})$
and 
\begin{equation}
h_{0}=\left(\begin{array}{cccc}
\varepsilon_{1} & 0 & t_{d} & 0\\
0 & -\varepsilon_{1} & 0 & -t_{d}\\
t_{d} & 0 & \varepsilon_{2} & 0\\
0 & -t_{d} & 0 & -\varepsilon_{2}
\end{array}\right),
\end{equation}
and
\begin{equation}
h_{1}=\left(\begin{array}{cccc}
0 & 0 & 0 & 0\\
0 & 0 & 0 & 0\\
0 & 0 & 0 & 1\\
0 & 0 & 1 & 0
\end{array}\right).
\end{equation}
We then define four fermionic operators $\hat{\gamma}_{i,\sigma}$
with $i=1,2$ and $\sigma=\uparrow,\downarrow$ such that $\hat{U}^{\dagger}h_{0}\hat{U}$
is diagonal, where $\hat{U}^{\dagger}=(\hat{\gamma}_{1,\uparrow}^{\dagger},\hat{\gamma}_{1,\downarrow},\hat{\gamma}_{2,\uparrow}^{\dagger},\hat{\gamma}_{2,\downarrow})$.
This is possible when the operators $\hat{\gamma}_{i,\sigma}$ are
related to $\hat{d}_{i,\sigma}$ operators through a Bogoliubov transformation
$\hat{U}=K\hat{\Psi}$, where $K$ is given by 
\begin{equation}
K=\left(\begin{array}{cccc}
u & 0 & v & 0\\
0 & u & 0 & v\\
-v & 0 & u & 0\\
0 & -v & 0 & u
\end{array}\right),\label{eq:eq14}
\end{equation}
where 
\begin{equation}
u(v)=\frac{1}{\sqrt{2}}\sqrt{1\pm\frac{\varepsilon_{L}-\varepsilon_{R}}{\sqrt{(\varepsilon_{L}-\varepsilon_{R})^{2}+4t_{d}^{2}}}}.\label{eq:eq15}
\end{equation}
Then, we find that $\hat{U}^{\dagger}h_{0}\hat{U}=\textrm{diag}(E_{1},-E_{1},E_{2},-E_{2})$,
where $E_{1}(E_{2})=\frac{1}{2}(\varepsilon_{L}+\varepsilon_{R}\pm\sqrt{(\varepsilon_{L}-\varepsilon_{R})^{2}+4t_{d}^{2}})$. 

To take into account the effect of the second term of Eq.(\ref{eq:eq12}),
we perform a first order perturbation on the eigenfunctions of the
matrix $h_{0}$ with respect to the perturbation term $\Gamma_{S}h_{1}$.
These perturbed eigenfunctions can be represented by 
\begin{equation}
\hat{U}=(K+\Gamma_{S}K^{\prime})\hat{\Psi},\label{eq:eq17-1}
\end{equation}
where 
\begin{gather}
K^{\prime}=\hspace{7cm}\nonumber \\
\left(\begin{array}{cccc}
{\scriptstyle 0} & {\scriptstyle \frac{uv^{2}}{2E_{1}}-\frac{uv^{2}}{E_{1}+E_{2}}} & {\scriptstyle 0} & {\scriptstyle \frac{v^{3}}{2E_{1}}+\frac{u^{2}v}{E_{1}+E_{2}}}\\
{\scriptstyle -\frac{uv^{2}}{2E_{1}}+\frac{uv^{2}}{E_{1}+E_{2}}} & {\scriptstyle 0} & {\scriptstyle -\frac{v^{3}}{2E_{1}}-\frac{u^{2}v}{E_{1}+E_{2}}} & {\scriptstyle 0}\\
{\scriptstyle 0} & {\scriptstyle \frac{u^{2}v}{E_{2}+E_{1}}-\frac{u^{2}v}{2E_{2}}} & {\scriptstyle 0} & {\scriptstyle \frac{uv^{2}}{E_{2}+E_{1}}+\frac{u^{3}}{2E_{2}}}\\
{\scriptstyle -\frac{u^{2}v}{E_{2}+E_{1}}+\frac{u^{2}v}{2E_{2}}} & {\scriptstyle 0} & {\scriptstyle -\frac{uv^{2}}{E_{2}+E_{1}}-\frac{u^{3}}{2E_{2}}} & {\scriptstyle 0}
\end{array}\right).\label{eq:eq17}
\end{gather}
Now, it is straightforward to show by matrix multiplication that the
first order correction to the eigenenergies of $\hat{H}_{dqd}^{SC}$
is zero and its diagonalized form in terms of $\hat{\gamma}_{i,\sigma}$
operators can be given by 
\begin{equation}
\hat{\widetilde{H}}_{dqd}^{SC}=U^{\dagger}\hat{H}_{dqd}^{SC}U=\sum_{i,\sigma}E_{i}\hat{\gamma}_{i,\sigma}^{\dagger}\hat{\gamma}_{i,\sigma}+\mathcal{O}(\Gamma_{S}^{2}).
\end{equation}

The main benefit which we can obtain from above calculations is that
we can apply the same unitary transformation $U$ on the DQD-resonator
coupling Hamiltonian $\hat{H}_{ph-dqd}$ to obtain its representation
in the terms of $\hat{\gamma}_{i,\sigma}$ operators as:
\begin{gather}
\hat{\widetilde{H}}_{ph-dqd}=-\sum_{\sigma}(\hat{a}+\hat{a}^{\dagger})[(g_{L}-g_{R})\hspace{2cm}\nonumber \\
\times\left[-uv\hat{\gamma}_{1,\sigma}^{\dagger}\hat{\gamma}_{2,\sigma}+\Gamma_{S}uv\left(\frac{u^{2}}{2E_{2}}+\frac{v^{2}}{2E_{1}}\right)\hat{\gamma}_{1,\sigma}^{\dagger}\hat{\gamma}_{2,-\sigma}^{\dagger}\right]\nonumber \\
\hspace{1cm}-(g_{L}+g_{R})\left[\Gamma_{S}\frac{uv}{E_{1}+E_{2}}\hat{\gamma}_{1,\sigma}^{\dagger}\hat{\gamma}_{2,-\sigma}^{\dagger}\right]]+h.c.
\end{gather}

\section{\label{sec:Lehmann}Lehmann representation of $G_{i}^{R}(\omega)$ }

As we discussed in Sec.\ref{ III}, to the linear order of interaction
between the DQD and the resonator and in the weak coupling regime,
it is reasonable to expect that the possible lasing frequencies in
the resonator can be explained by electron transitions between various
peaks in the LDOS of the DQD when it is isolated from the resonator
and also from the normal lead. It is convenient to to calculate the
LDOS of the DQD from $-\textrm{Im}[G_{i}^{R}(\omega)]$, where $G_{i}^{R}(\omega)$
is the retarded Green's function of the dot $i$ and can be calculated
from the Lehmann representation.\citep{bruus2004many} In the following
we briefly summarize the main steps needed to calculate the retarded
Green's function of the DQD.

Our starting point is the Hamiltonian of the DQD 
\begin{align}
\hat{H}_{dqd}^{SC}= & \sum_{\alpha\sigma}\varepsilon_{\alpha}\hat{n}_{\alpha,\sigma}+U_{\alpha}\hat{n}_{\alpha,\uparrow}\hat{n}_{\alpha,\downarrow}+U_{LR}\hat{n}_{L}\hat{n}_{R}\nonumber \\
 & +t_{d}\sum_{\sigma}(d_{R,\sigma}^{\dagger}d_{L,\sigma}+h.c)+\Gamma_{S}(\hat{d}_{R,\uparrow}^{\dagger}\hat{d}_{R,\downarrow}^{\dagger}+h.c).\label{eq:HDQD}
\end{align}
To proceed, we need the eigenstates and the eigenenergies of Eq.(\ref{eq:HDQD}),
which as we discussed in Sec.\ref{II}, obtaining an analytical expression
for its eigenstates is practically impossible. Nevertheless, we can
rely on numerical methods and express the numerically calculated eigenstates
of Eq.(\ref{eq:HDQD}) and their corresponding eigenenergies by $\bigl|n\bigr\rangle$
and $\mathcal{E}_{n}$ where $n=1,2,\cdots,16$. Having the eigenspectrum
of the Hamiltonian of the DQD, we can then calculate the retarded
Green's function of the DQD from
\begin{equation}
G_{i,\sigma}^{R}(\omega)=\frac{1}{Z}\sum_{m,m^{\prime}}\frac{\bigl\langle m\bigr|d_{i,\sigma}\bigl|m^{\prime}\bigr\rangle\bigl\langle m^{\prime}\bigr|d_{i,\sigma}^{\dagger}\bigl|m\bigr\rangle}{\omega+\mathcal{E}_{m}-\mathcal{E}_{m^{\prime}}+i\eta}\left(e^{-\beta\mathcal{E}_{m}}+e^{-\beta\mathcal{E}_{m^{\prime}}}\right),
\end{equation}
where $Z=\sum_{m}e^{-\beta\mathcal{E}_{m}}$, $\eta$ is an infinitesimal
positive constant and $\beta$ is the inverse temperature. 

\bibliographystyle{apsrev4-1}
\bibliography{ref_lasing}

\end{document}